\documentclass[twocolumn,showpacs,preprintnumbers,amsmath,amssymb,superscriptaddress,nofootinbib,english]{revtex4-1}
\pdfoutput=1
\usepackage{times,amsmath,amsfonts,amssymb,epstopdf}
\usepackage{graphicx}
\usepackage{dcolumn}
\usepackage{bm}
\usepackage{epsfig}
\usepackage{graphicx}
\usepackage{hyperref}
\usepackage[usenames]{color}
\usepackage{url}
\usepackage[normalem]{ulem}
\usepackage[T1]{fontenc}

\def\be{\begin{equation}}
\def\ee{\end{equation}}
\def\ben{\begin{eqnarray}}
\def\een{\end{eqnarray}}
\def\ba{\begin{array}}
\def\ea{\end{array}}

\newcommand{\bq}{\begin{eqnarray}}
\newcommand{\eq}{\end{eqnarray}}
\newcommand{\bes}{\begin{subequations}}
\newcommand{\ees}{\end{subequations}}

\begin{document}
\newcommand{\half}{{\textstyle\frac{1}{2}}}
\allowdisplaybreaks[3]
\def\triangledown{\nabla}
\def\grad3{\hat{\nabla}}
\def\a{\alpha}
\def\b{\beta}
\def\g{\gamma}\def\G{\Gamma}
\def\d{\delta}\def\D{\Delta}
\def\ep{\epsilon}
\def\et{\eta}
\def\z{\zeta}
\def\t{\theta}\def\T{\Theta}
\def\l{\lambda}\def\L{\Lambda}
\def\m{\mu}
\def\f{\phi}\def\F{\Phi}
\def\n{\nu}
\def\p{\psi}\def\P{\Psi}
\def\r{\rho}
\def\s{\sigma}\def\S{\Sigma}
\def\ta{\tau}
\def\x{\chi}
\def\o{\omega}\def\O{\Omega}
\def\k{\kappa}
\def\pa {\partial}
\def\ov{\over}
\def\br{\\}
\def\ud{\underline}

\newcommand\lsim{\mathrel{\rlap{\lower4pt\hbox{\hskip1pt$\sim$}}
    \raise1pt\hbox{$<$}}}
\newcommand\gsim{\mathrel{\rlap{\lower4pt\hbox{\hskip1pt$\sim$}}
    \raise1pt\hbox{$>$}}}
\newcommand\esim{\mathrel{\rlap{\raise2pt\hbox{\hskip0pt$\sim$}}
    \lower1pt\hbox{$-$}}}
\newcommand{\dpar}[2]{\frac{\partial #1}{\partial #2}}
\newcommand{\sdp}[2]{\frac{\partial ^2 #1}{\partial #2 ^2}}
\newcommand{\dtot}[2]{\frac{d #1}{d #2}}
\newcommand{\sdt}[2]{\frac{d ^2 #1}{d #2 ^2}}    

\preprint{IPPP/13/36}
\preprint{DCPT/13/72}

\title{Nonlinear structure formation in the Cubic Galileon gravity model}

\author{Alexandre Barreira}
\email[Electronic address: ]{a.m.r.barreira@durham.ac.uk}
\affiliation{Institute for Computational Cosmology, Department of Physics, Durham University, Durham DH1 3LE, U.K.}
\affiliation{Institute for Particle Physics Phenomenology, Department of Physics, Durham University, Durham DH1 3LE, U.K.}

\author{Baojiu Li}
\affiliation{Institute for Computational Cosmology, Department of Physics, Durham University, Durham DH1 3LE, U.K.}

\author{Wojciech A. Hellwing}
\affiliation{Institute for Computational Cosmology, Department of Physics, Durham University, Durham DH1 3LE, U.K.}
\affiliation{Interdisciplinary Centre for Mathematical and Computational Modeling (ICM), University of Warsaw, ul. Pawi\'nskiego 5a, Warsaw, Poland}

\author{Carlton M. Baugh}
\affiliation{Institute for Computational Cosmology, Department of Physics, Durham University, Durham DH1 3LE, U.K.}

\author{Silvia Pascoli}
\affiliation{Institute for Particle Physics Phenomenology, Department of Physics, Durham University, Durham DH1 3LE, U.K.}

\begin{abstract}

We model the linear and nonlinear growth of large scale structure in the Cubic Galileon gravity model, by running a suite of N-body cosmological simulations using the {\tt ECOSMOG} code. Our simulations include the Vainshtein screening effect, which reconciles the Cubic Galileon model with local tests of gravity. In the linear regime, the amplitude of the matter power spectrum increases by $\sim 20\%$ with respect to the standard $\Lambda$CDM model today. The modified expansion rate accounts for $\sim 15\%$ of this enhancement, while the fifth force is responsible for only $\sim 5\%$. This is because the effective unscreened gravitational strength deviates from standard gravity only at late times, even though it can be twice as large today. In the nonlinear regime ($k \gtrsim 0.1 h\rm{Mpc}^{-1}$), the fifth force leads to only a modest increase ($\lesssim 8\%$) in the clustering power on all scales due to the very efficient operation of the Vainshtein mechanism. Such a strong effect is typically not seen in other models with the same screening mechanism. The screening also results in the fifth force increasing the number density of halos by less than $10\%$, on all mass scales.  Our results show that the screening does not ruin the validity of linear theory on large scales which anticipates very strong constraints from galaxy clustering data. We also show that, whilst the model gives an excellent match to CMB data on small angular scales ($l \gtrsim 50$), the predicted integrated Sachs-Wolfe effect is in tension with Planck/WMAP results.

\end{abstract} 
%\pacs{98.80.Cq}
\maketitle

\section{Introduction}

Measurements of the cosmic microwave background (CMB) \cite{Hinshaw:2012fq, Ade:2013zuv}, type Ia supernovae (SNIa) \cite{Guy:2010bc, Suzuki:2011hu}, Hubble constant \cite{Riess:2009pu, Riess:2011yx, Freedman:2012ny}, galaxy clustering and the scale of the baryonic acoustic oscillation (BAO) feature \cite{Efstathiou:2001cw, Sanchez:2009jq, Percival:2009xn, Beutler:2011hx, Reid:2012sw, Anderson:2012sa, Sanchez:2012sg} now provide compelling evidence that our Universe is going through a period of accelerated expansion. This suggests the existence of some form of  "dark energy", the fundamental nature of which currently evades our understanding. Nevertheless, there is no shortage of possible explanations (see \cite{2006IJMPD..15.1753C, Li:2011sd, Clifton:2011jh} for recent reviews), with the best known and simplest being the cosmological constant $\Lambda$. In the so called $\Lambda$-cold dark matter ($\Lambda$CDM) model, $\Lambda$ "inflates" the universe due to its negative pressure. The problem with this model, however, is that its recognized observational success comes at the expense of requiring a heavily fine-tuned value of $\Lambda$, which is hard to motivate theoretically. 

An alternative to negative-pressure dark energy is to modify the standard gravitational law of general relativity (GR). A recent prominent example is the case of Galileon gravity \cite{PhysRevD.79.064036, PhysRevD.79.084003, Deffayet:2009mn}, in which the modifications to GR arise through a Galilean-invariant scalar field, i.e., it is invariant under the transformation $\partial_\mu\varphi \rightarrow \partial_\mu\varphi + b_\mu$, where $b_\mu$ is a constant vector. The action of the Galileon model was first derived in \cite{PhysRevD.79.064036}, by generalizing the four-dimensional effective action of the Dvali-Gabadadze-Porrati (DGP) model \cite{2000PhLB..485..208D, 1126-6708-2003-09-029, 1126-6708-2004-06-059, deRham:2012az}. The latter already had a Galilean degree of freedom which was, however, plagued by ghost problems \cite{1126-6708-2003-09-029, 1126-6708-2004-06-059, PhysRevD.73.044016}. The authors of \cite{PhysRevD.79.064036} found that in four-dimensional Minkowski space, one can only write five Lagrangian densities that are invariant under the Galilean-shift transformation and which lead to second-order field equations of motion, despite nonlinear derivative terms that appear in the Lagrangian. The second-order nature of the equations makes the Galileon model a subset of the more general Horndeski theory \cite{Horndeski:1974wa}, and is crucial to avoid the propagation of Ostrogradsky ghosts \cite{Woodard:2006nt}. The action of \cite{PhysRevD.79.064036} was subsequently generalized to curved spacetimes in \cite{PhysRevD.79.084003, Deffayet:2009mn}, where it was found necessary to explicitly couple the Galileon derivatives to curvature tensors. Such terms make the modifications of gravity highly non-trivial.

The main difficulty in modified gravity theories is to reconcile $\mathcal{O}(1)$ modifications on cosmological scales with local tests of gravity. The former are needed to accelerate the expansion of the Universe, while the latter constrain gravity to be very close to GR on small scales \cite{1990PhRvL..64..123D, 1999PhRvL..83.3585B, Will:2005va, Kapner:2006si}. Consequently, a crucial requirement of these models is that they should possess some form of screening mechanism that suppresses the modifications to gravity\footnote{Often described as a fifth force.} on the scales where GR is well tested. In the case of the Galileon model, the screening is realized by the nonlinear derivative terms of the Lagrangian through an effect known as the Vainshtein mechanism \cite{Vainshtein1972393, Babichev:2013usa, Koyama:2013paa}. In this case, nonlinearities result in interference terms that weaken the modifications to GR in regions where the matter density is higher, such as the Solar System. In particular, near matter sources, nonlinear terms become important and suppress the spatial gradient of the Galileon field, which corresponds to the fifth force. Far away from the source, the nonlinear interference is weak and the fifth force contributes substantially to the total force felt by a test particle. The DGP model, massive gravity \cite{deRham:2010kj, Sbisa:2012zk, deRham:2011by}, Fab-Four \cite{Charmousis:2011bf, Charmousis:2011ea, Bruneton:2012zk, Copeland:2012qf} and kinetic gravity braiding \cite{Deffayet:2010qz, Pujolas:2011he, Kimura:2011td}, are examples of other models where the Vainshtein mechanism can be implemented (see also \cite{Appleby:2012rx, Babichev:2009ee, Kobayashi:2009wr, Kobayashi:2010wa, Leon:2012mt, Silva:2009km}). Another way to screen the fifth force in modified gravity theories is via nonlinear interactions of a scalar field with matter and/or itself. This mechanism, widely known as the "chameleon" effect, works by choosing the nonlinear matter coupling and potential functions of the models in such a way that, in high-density environments, they lead to vanishing coupling strengths (e.g. dilaton \cite{Brax:2010gi, Brax:2011aw, Brax:2012gr} and symmetron \cite{Hinterbichler:2011ca, Hinterbichler:2010es, Davis:2011pj, Brax:2011pk} models) and/or to very heavy masses of the scalar field (e.g. Chameleon scalar fields \cite{PhysRevD.69.044026, PhysRevD.70.123518} and $f(R)$ \cite{Sotiriou:2008rp, DeFelice:2010aj, Carroll:2004de, Brax:2008hh, Li:2007xn, Hu:2007nk}). The heavier the scalar field, the harder it is for it to propagate, and therefore, to mediate any fifth force. Recently, another type of screening mechanism based on disformal couplings has also been proposed \cite{Koivisto:2012za}.

The cosmological properties of the Galileon gravity model have recently been studied at the linear level in perturbation theory \cite{Barreira:2012kk, Barreira:2013jma, Gannouji:2010au, PhysRevD.80.024037, DeFelice:2010pv, Nesseris:2010pc, Appleby:2011aa, PhysRevD.82.103015, Neveu:2013mfa, Appleby:2012ba, Okada:2012mn, Bartolo:2013ws}. In particular, in \cite{Barreira:2012kk}, we have modified the Boltzmann code {\tt CAMB} \cite{camb_notes} to perform the first study of the CMB temperature and linear matter power spectra in the Galileon model.  As a follow up of this study, in \cite{Barreira:2013jma} we carried out a formal Markov Chain Monte Carlo exploration of the full cosmological parameter space in the general covariant Galileon gravity model of \cite{PhysRevD.79.084003}. We have found that the Galileon model can provide a reasonable fit to the current data for the CMB, SNIa and BAO, and in fact, is a better fit than $\Lambda$CDM if only the CMB data is considered. However, we also have identified some tension in the ability of the Galileon model to fit both the CMB and galaxy clustering data (see also \cite{Appleby:2012ba, Okada:2012mn, Neveu:2013mfa}). In principle, such tension could be used to place very strong constraints on the model, perhaps even ruling it out. However, such a conclusion depends critically on the assumed validity of linear perturbation theory on the scales probed by the data, which is less obvious in the case of modified gravity models due to the presence of the nonlinear screening effects. The screening can have a substantial impact on scales where linear theory is usually taken to be a good approximation, as shown in \cite{Oyaizu:2008tb, Brax:2012nk, Brax:2011ja, Jennings:2012pt, Li:2012by, Brax:2013mua, Hellwing:2013rxa}, via N-body simulations in the case of chameleon-type models.

Hence, in order to properly constrain Galileon gravity, it is important to understand the true impact of the Vainshtein screening on the model, which requires one to go beyond linear theory. Analytical studies of nonlinear structure formation are usually limited to special configurations, such as assuming spherical symmetry \cite{Bellini:2012qn, Kimura:2011dc, Babichev:2011iz, DeFelice:2011th, Hiramatsu:2012xj, Burrage:2010rs, Iorio:2012pv, Brax:2011sv, deRham:2012fw, deRham:2012fg, Garcia-Saenz:2013gya, Berezhiani:2013dw, Berezhiani:2013dca, Ali:2012cv}. Therefore, one has to resort to N-body simulations to study the nonlinearities at a cosmological level. This is not trivial for models which employ Vainshtein screening, as the relevant equations are difficult to solve numerically due to the products of derivative terms. This is particularly challenging in the case of the Galileon model, where one can have terms with products of five field derivatives. In this paper, as a first step to understanding the influence of the Vainshtein screening on nonlinear structure formation in a Galileon cosmology, we consider the special case of the Cubic Galileon model. This is a theoretically consistent subset of the general theory of \cite{PhysRevD.79.064036, PhysRevD.79.084003, Deffayet:2009mn}, but in which the degree of nonlinearity in the equations of motion is smaller. This makes the equations easier to solve numerically.

The Cubic Galileon model is, in many aspects, similar to the DGP model, of which a number of N-body studies have already been performed \cite{Schmidt:2009sg, Schmidt:2009sv, Chan:2009ew, Khoury:2009tk, Schmidt:2009yj, Li:2013nua}. An important advantage of the Cubic Galileon model is that it fits the observational data much better than the DGP model, and that it can have a self-accelerated solution without the presence of ghosts \cite{1126-6708-2003-09-029, 1126-6708-2004-06-059, PhysRevD.73.044016}. In \cite{Wyman:2013jaa}, the authors have simulated a phenomenological model with equations similar to the DGP and Cubic Galileon models, but by making a number of assumptions regarding the time evolution of the expansion rate and the fifth force. In particular, the model of \cite{Wyman:2013jaa} assumes a $\Lambda$CDM expansion history, which is not possible in the Galileon model. Moreover, the evolution of the fifth force is also different, which, as we will see below, plays an important role in the growth of structure in the two models. In \cite{Li:2013nua}, the authors reported an upgrade of the {\tt ECOSMOG} N-body code \cite{Li:2011vk} to perform high-resolution simulations which include Vainshtein screening, by taking advantage of the adaptive mesh refinement (AMR) nature of the code. In this paper, we use the same N-body code to look at how the Cubic Galileon model compares to the models simulated in \cite{Li:2013nua, Wyman:2013jaa}, and what new phenomena could arise.

The outline of the paper is as follows. In Sec. \ref{model-section}, we present the relevant equations for the N-body simulations, derived under the quasi-static limit approximation. We also look at the implementation of the Vainshtein mechanism, the CMB power spectrum of the model and its linear density field evolution. In Sec. \ref{simulations-section}, we briefly describe the N-body algorithm and the simulations we have performed. The results for the nonlinear matter power spectrum and halo mass function are presented in Sec.~\ref{results-section}. We conclude in Sec. \ref{conclusion-section}.

Throughout this paper we assume the metric convention $(+,-,-,-)$ and work in units in which the speed of light $c = 1$. Greek indices run over $0,1,2,3$ and we use $8\pi G=\kappa=M^{-2}_{\rm Pl}$ interchangeably, where $G$ is Newton's constant and $M_{\rm Pl}$ is the reduced Planck mass.

\section{The Galileon model}\label{model-section}

\label{The model}

Here, we present the Cubic Galileon model and the equations that are relevant for the formation of large-scale structure. A goal of this section is to contrast the Cubic Galileon model with the DGP braneworld model. We follow a similar notation to that adopted in \cite{Li:2013nua}, to which we refer the reader for all the relevant DGP expressions.

\subsection{Action and field equations}\label{equations}

The action of the Cubic Galileon model, which has no direct coupling between matter and the Galileon field (see however \cite{Appleby:2011aa,  Sushkov:2009hk, Gubitosi:2011sg, deRham:2011by, VanAcoleyen:2011mj, Zumalacarregui:2012us, Amendola:1993uh, Barreira:2012kk}), is given by

\bq\label{Galileon action}
&& S = \int d^4x\sqrt{-g} \left[ \frac{R}{16\pi G} - \frac{1}{2}c_2\mathcal{L}_2 - \frac{1}{2}c_3\mathcal{L}_3 - \mathcal{L}_m\right],
\eq
where $g$ is the determinant of the metric $g_{\mu\nu}$, $R$ is the Ricci scalar, the model parameters $c_2$ and $c_3$ are dimensionless constants and $\mathcal{L}_2$ and $\mathcal{L}_3$ are given by

\bq\label{L's}
\mathcal{L}_2 = \nabla_\mu\varphi\nabla^\mu\varphi,\ \ \ \ \ \ \ \  \mathcal{L}_3 = \frac{2}{M^3}\Box\varphi\nabla_\mu\varphi\nabla^\mu\varphi,
\eq
in which $\varphi$ is the Galileon scalar field and $\mathcal{M}^3\equiv M_{\rm Pl}H_0^2$, where $H_0$ is the present-day Hubble expansion rate.  

The Einstein and Galileon field equations are given by:

\bq
\label{full-Einstein}G_{\mu\nu} &=& \kappa \left[T_{\mu\nu}^m + T_{\mu\nu}^{c_2} + T_{\mu\nu}^{c_3}\right], \\
\label{full-EoM}0 &=& c_2\Box\varphi + 2\frac{c_3}{\mathcal{M}^3}\left[(\Box\varphi)^2 - \nabla^\alpha\nabla^\beta\varphi\nabla_\alpha\nabla_\beta\varphi \right. \nonumber \\
 && \ \ \ \ \ \ \ \ \ \ \ \ \ \ \ \ \ \ \ \ \ \ \ \ \ \ \left. - R_{\alpha\beta}\nabla^\alpha\varphi\nabla^\beta\varphi \right],
\eq
where 
\bq
T_{\mu\nu}^{c_2} &=& c_2\left[ \nabla_\mu\varphi\nabla_\nu\varphi - \frac{1}{2}g_{\mu\nu}\nabla^\alpha\varphi\nabla_\alpha\varphi \right], \\ 
T_{\mu\nu}^{c_3} &=& \frac{c_3}{\mathcal{M}^3}\left[ 2\nabla_\mu\varphi\nabla_\nu\varphi\Box\varphi + 2g_{\mu\nu}\nabla_\alpha\varphi\nabla_\beta\varphi\nabla^\alpha\nabla^\beta\varphi \right. \nonumber \\
&& \ \ \ \ \ \ \ \ \ \left. - 4\nabla^\lambda\varphi\nabla_{(\mu}\varphi\nabla_{\nu)}\nabla_\lambda\varphi \right],
\eq
and $T_{\mu\nu}^m$ is the energy-momentum tensor of all the other energy species in the universe.

Before proceeding, note that the action of the Cubic Galileon (Eq.~(\ref{Galileon action})) represents only a sector of the more general Galileon gravity model \cite{PhysRevD.79.064036, PhysRevD.79.084003, Deffayet:2009mn}. The latter contains two more Lagrangian densities that are quartic and quintic in derivatives of the scalar field. Nevertheless, as pointed out in \cite{PhysRevD.79.064036, Burrage:2012ja}, the cubic subset suffers from less serious theoretical problems related to the smallness of the energy cutoff below which the phenomenological theory is valid. The general Galileon model contains also a linear potential term, which is often neglected if one is interested only in cases where cosmic acceleration is driven by field kinetic terms. 

\subsubsection{Background equations}

We will work with the perturbed Friedmann-Robertson-Walker (FRW) metric in the Newtonian gauge

\bq\label{metric}
ds^2 = \left(1 + 2\Psi\right)dt^2 - a(t)^2\left(1 - 2\Phi\right)\gamma_{ij}dx^idx^j,
\eq
where $\gamma_{ij} = \rm{diag}\left[1, 1, 1\right]$ is the spatial sector of the metric, which is taken here to be flat. The Galileon field, $\varphi$, as well as the scalar potentials, $\Psi$, $\Phi$, are assumed to be functions of time and space.  In the equations below $\varphi = \bar{\varphi}(t) + \delta\varphi(t, \vec{x})$, where $\delta\varphi$ is the field perturbation and an overbar denotes background averaged quantities. Note we shall always use $\varphi$, and the context should determine whether we refer to $\bar{\varphi}$ or $\delta\varphi$.

At the cosmological background level ($\delta\varphi = \Phi = \Psi = 0$), the two Friedmann equations are given by

\bq
\label{Friedman1}3H^2 &=& \kappa\left[\bar{\rho}_m + \frac{c_2}{2}\dot{\varphi}^2 + 6\frac{c_3}{\mathcal{M}^3} \dot{\varphi}^3H \right], \\
\label{Friedman2}\dot{H} + H^2 &=& -\frac{\kappa}{6}\left[ \bar{\rho}_m + 4c_2\dot{\varphi}^2 \right. \nonumber \\
&& \ \ \ \ \ \ \ \ \left. + 6\frac{c_3}{\mathcal{M}^3} \left(\dot{\varphi}^3H - \ddot{\varphi}\dot{\varphi}^2 \right) \right],
\eq
and the Galileon field equation of motion by

\bq\label{background-EoM}
0 &=& c_2\left[\ddot{\varphi} +3\dot{\varphi}H \right] \nonumber \\
&& + 6\frac{c_3}{\mathcal{M}^3} \left[ 2\ddot{\varphi}\dot{\varphi}H + 3\dot{\varphi}^2H^2 + \dot{\varphi}^2\dot{H}\right],
\eq
where $\rho_m$ is the matter density and we have omitted the contribution from radiation since this is negligible during the late times probed by our simulations. Eqs.~(\ref{Friedman1}), (\ref{Friedman2}) and (\ref{background-EoM}) govern the background expansion history of the univere, which determines how the distances between the particles in the simulation change with time.

\begin{figure}
	\centering
	\includegraphics[scale=0.49]{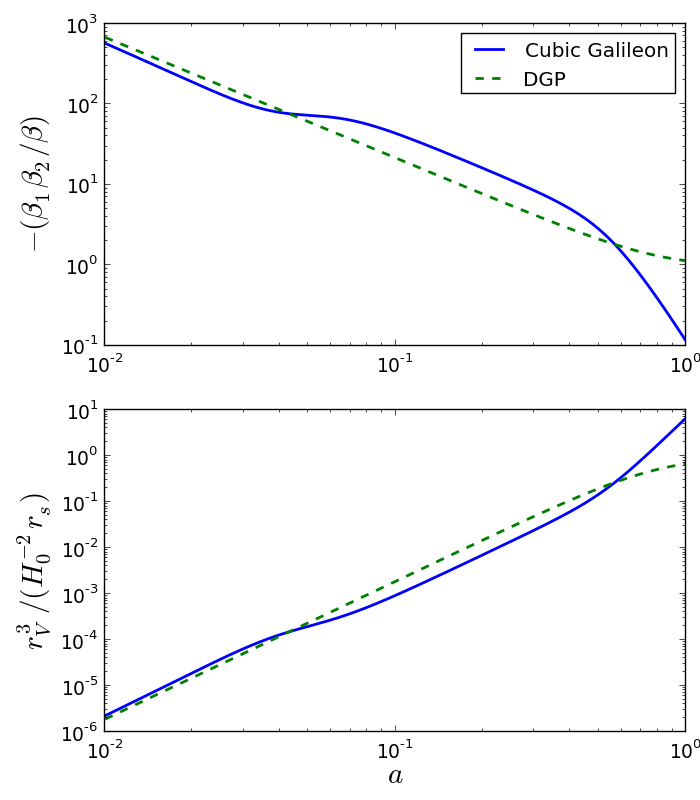}
	\caption{Time evolution of the quantity $-\beta_1\beta_2/\beta$, with $\beta=\beta_{DGP}$ (c.f.~Eqs.~(\ref{beta1}), (\ref{beta2}) and (\ref{poisson+eom-2})) (top panel), and of the Vainstein radius $r_V$ (bottom panel) for the Cubic Galileon (solid blue) and DGP (dashed green) models. In the bottom panel, we have assumed $r_c = H_0^{-1}$ so that $r_V$ can be plotted in the same units for both models. The Cubic Galileon model plotted is the model of Table \ref{table-max} while the DGP model is the model simulated in \cite{Li:2013nua}.}
\label{beta-rv}\end{figure}

\subsubsection{Quasi-static approximation}\label{quasi-static-section}

The quasi-static limit is known to be a good approximation in the Galileon model for the sub-horizon scales probed by our simulations \cite{Barreira:2012kk, DeFelice:2010as}. This is the limit in which the time derivatives of the perturbed quantities can be neglected compared with their spatial derivatives. Additionally, to further simplify the equations, we can also neglect terms that are suppressed by the Newtonian potentials, $\Phi$ and $\Psi$, and their first spatial derivatives, $\Phi,_i$ and $\Psi,_i$, since these quantities are typically very small ($\lesssim 10^{-4}$) on the scales of interest. For instance, $\left(1-2\Phi\right)\partial^i\partial_i\varphi \sim \partial^i\partial_i\varphi$ or $\partial_i\partial^i\Phi\partial_j\Phi \ll \partial_i\partial^i\Phi$. The validity of these assumptions can always be assessed by checking if the simulation results reproduce the full linear theory predictions on the scales where linear theory should hold. In our notation, $\partial_i$ denotes a partial derivative w.r.t.~the $i$-th spatial coordinate ($i = x,y,z$), and the indices are lowered and raised using the spatial metric $\gamma_{ij}$ and its inverse $\gamma^{ij}$, respectively.

Under the above approximations, the Poisson equation (obtained via the $(0,0)$ component of Eq.~(\ref{full-Einstein})) and the Galileon field equation of motion, Eq.~(\ref{full-EoM}), are given, respectively, by

\bq
&&\label{poisson}\partial^2\Phi = 4\pi Ga^2\delta\rho_m - \frac{\kappa c_3}{\mathcal{M}^3}\dot{\varphi}^2\partial^2\varphi, \\
&&\label{perturbed-EoM}\frac{2c_3}{\mathcal{M}^3}\dot{\varphi}^2\partial^2\Psi = \left[-c_2 - \frac{4c_3}{\mathcal{M}^3}\left(\ddot{\varphi} + 2H\dot{\varphi}\right)\right]\partial^2\varphi \nonumber \\
&&\ \ \ \ \ \ \ \ \ \ \ \ \ \ \ \ \ + \frac{2c_3}{a^2\mathcal{M}^3}\left[(\partial^2\varphi)^2 - (\partial_i\partial_ j\varphi)^2\right].
\eq
Eqs.~(\ref{poisson}) and (\ref{perturbed-EoM}) can be combined in an equation that involves solely the Galileon field and the matter density perturbation:

\bq\label{poisson+eom}
&&\partial^2{\varphi} + \frac{1}{3\beta_1a^2\mathcal{M}^3}\left[(\partial^2{\varphi})^2 - (\partial_i\partial_ j{\varphi})^2\right]  = \nonumber \\
&& \ \ \ \ \ \ \ \ \ \ \ \ \ \ \ \ \ \ \ \ \ \ \ \ \ \ \ \ \ \ \ \ \ \ \ \ \ = \frac{M_{\rm{Pl}}}{3\beta_2}8\pi Ga^2\delta\rho_m,
\eq
where we have used the relation $\Phi = \Psi$ in the Cubic Galileon model, as a consequence of the vanishing anisotropic stress  \cite{Barreira:2012kk}. Here, $\partial^2 = \partial_i\partial^i$ is the spatial Laplacian differential operator and  $(\partial_i\partial_ j{\varphi})^2 = (\partial_i\partial_ j{\varphi})(\partial^i\partial^j{\varphi})$. $\delta\rho_m$ is the matter density perturbation, $\rho_m = \bar{\rho}_m(t) + \delta\rho_m(t, \vec{x})$. The dimensionless functions $\beta_1$ and $\beta_2$ are defined as

\bq
\label{beta1}\beta_1 &=&\frac{1}{6c_3} \left [-c_2 - \frac{4c_3}{\mathcal{M}^3}\left(\ddot{\varphi} + 2H\dot{\varphi}\right) + 2\frac{\kappa c_3^2}{\mathcal{M}^6}\dot{\varphi}^4\right], \\
\label{beta2}\beta_2 &=& 2\frac{\mathcal{M}^3M_{\rm{Pl}}}{\dot{\varphi}^2}\beta_1.
\eq

Eq.~(\ref{poisson+eom}) has the same structural form (in terms of the spatial derivatives of the scalar field) as the equation of motion of the DGP brane-bending mode \cite{Koyama:2007ih}. The differences lie only in the distinct time evolution of the functions $\beta_1$ and $\beta_2$. In particular, in the DGP model $\beta_1 = \beta_2$.  To facilitate the comparison between the different models with equations of the same form as Eqs.~(\ref{poisson}) and (\ref{poisson+eom}), we can redefine the field perturbation as 

\bq\label{field-perturbation-redefinition}
\delta\varphi \rightarrow \frac{\beta}{\beta_2}\delta\varphi,
\eq
where $\beta$ is a free function. With this redefinition, Eqs.~(\ref{poisson}) and (\ref{poisson+eom}) become

\bq
&&\label{poisson-2}\partial^2\Phi = 4\pi Ga^2\delta\rho_m - \frac{\kappa c_3 }{\mathcal{M}^3}\frac{\beta}{\beta_2}\dot{\varphi}^2\partial^2\varphi, \\
&&\label{poisson+eom-2}\partial^2{\varphi} + \frac{1}{3\left(\beta_1\beta_2/\beta\right)a^2\mathcal{M}^3}\left[(\partial^2{\varphi})^2 - (\partial_i\partial_ j{\varphi})^2\right]  = \nonumber \\
&& \ \ \ \ \ \ \ \ \ \ \ \ \ \ \ \ \ \ \ \ \ \ \ \ \ \ \ \ \ \ \ \ \ \ \ \ \ = \frac{M_{\rm{Pl}}}{3\beta}8\pi Ga^2\delta\rho_m.
\eq
In this way, we can choose $\beta$ to make the right-hand side of Eq.~(\ref{poisson+eom-2}) look like in a given model, such as the DGP model. In this case, the differences between the two models in the scalar field equation are fully captured by the different values of $\beta_1\beta_2/\beta$ in the coefficient of the nonlinear derivative terms. In the top panel of Fig.~\ref{beta-rv}, we show the time evolution of $-\beta_1\beta_2/\beta$ for the Cubic Galileon (solid blue) and DGP (dashed green) models taking $\beta = \beta_{DGP}$. Here and throughout the paper, the DGP model is the self-accelerating branch that best fits the WMAP 5yr CMB data \cite{Fang:2008kc} and that has been simulated in \cite{Li:2013nua}. Note that in the DGP model, $\beta_1\beta_2/\beta_{DGP} = \beta_{DGP}$. We see that for both models the value of $-\beta_1\beta_2/\beta_{DGP}$ decreases overall but in different ways. In particular, the value of  $-\beta_1\beta_2/\beta_{DGP}$  in the Cubic Galileon model can be smaller or larger than in the DGP model throughout cosmic history. Since this term multiplies the nonlinear derivative terms, its different time evolution in these two models translates into a different efficiency for the screening mechanism, as we will see in the next section.

Note however, that besides the different coefficients of the nonlinear derivative terms, different models can also differ in the coefficient of $\partial^i\partial_i\varphi$ in the Poisson equation Eq.~(\ref{poisson-2}). In particular, in the Cubic Galileon model, such a coefficient is time-dependent whereas in the DGP model, for instance, it is simply a constant equal to $1/2$.

\subsection{Vainshtein screening}\label{vainshtein}

Equation (\ref{poisson+eom-2}) tells us that different models can be compared by the different coefficients of the nonlinear derivative term responsible for the screening. It is therefore instructive to understand how such derivative couplings work to suppress the modifications of gravity. For simplicity, here we look only at the case of spherically symmetric configurations of the gravitational and scalar fields. Therefore, assuming that $\varphi$ and $\Phi$ depend only on the radial coordinate, $r$, Eq.~(\ref{poisson+eom-2}) becomes:

\bq\label{spherical-eq}
&&\frac{1}{r^2}\frac{d}{dr}\left[r^2\varphi,_r\right] + \frac{2}{3}\frac{1}{\mathcal{M}^3a^2\left(\beta_1\beta_2/\beta\right)}\frac{1}{r^2}\frac{d}{dr}\left[r\varphi,_r^2\right] \nonumber \\
&& = \frac{M_{\rm{Pl}}}{3\beta}8\pi Ga^2 \delta\rho,
\eq
which can be integrated once to yield

\bq\label{spherical-eq-integrated}
\varphi,_r + \frac{2}{3}\frac{1}{\mathcal{M}^3a^2\left(\beta_1\beta_2/\beta\right)}\frac{1}{r}\varphi,_r^2 = \frac{2M_{\rm{Pl}}}{3\beta}\frac{GM(r)}{r^2}a^2,
\eq
where $M(r) = 4\pi\int_0^r\delta\rho_{m}(\xi)\xi^2d\xi$ is the matter contribution to the mass enclosed within a radius $r$. Eq.~(\ref{spherical-eq-integrated}) is a second-order algebraic equation for $\varphi,_r$. Taking for simplicity a top-hat density distribution of radius $R$, the physical solutions are given by:

\bq\label{algebraic-solution-1}
\varphi,_r = \frac{4M_{\rm{Pl}}a^2r^3}{3\beta r_V^3}\left[\sqrt{\left(\frac{r_V}{r}\right)^3 + 1} - 1\right]\frac{GM(R)}{r^2},
\eq
for $r \ge R$ and 

\bq\label{algebraic-solution-2}
\varphi,_r = \frac{4M_{\rm{Pl}}a^2R^3}{3\beta r_V^3}\left[\sqrt{\left(\frac{r_V}{R}\right)^3 + 1} - 1\right]\frac{GM(r)}{r^2},
\eq
for $r < R$. In Eqs.~(\ref{algebraic-solution-1}) and (\ref{algebraic-solution-2}) we have identified a distance scale, $r_V$, known as the Vainshtein radius, which is given by 

\bq\label{Vainshtein-radius}
r_V^3 = \frac{8M_{\rm{Pl}}r_S}{9M^3\beta_1\beta_2},
\eq
where $r_S \equiv 2GM(R)$ is the Schwarzschild radius of the top-hat source. The last term in the modified Poisson equation (Eq.~(\ref{poisson-2})) represents the fifth force mediated by the Galileon field:
\bq\label{fifth-force}
F_{5th} = -\frac{\kappa c_3}{\mathcal{M}^3}\frac{\beta}{\beta_2}\dot{\varphi}^2\varphi,_{r}.
\eq
Taking the limits where $r \gg r_V$ and $r \ll r_V$ one has

\bq
F_{5th} &=& -\frac{2c_3 a^2\dot{\varphi}^2}{3\mathcal{M}^3M_{\rm{Pl}}\beta_2}\frac{GM(R)}{r^2}\ \ \ \ r\gg r_V, \\
F_{5th} &\sim& 0 \ \ \ \ \ \ \ \ \ \ \ \ \ \ \ \ \ \ \ \ \ \ \ \ \ \ \ \ \ \ \ \ \ \ \ \ \ \ \ r \ll r_V.
\eq
Consequently, $r_V$ gives a measure of the length scale below which the screening mechanism starts to operate to recover the normal general relativistic force law. If $\beta_1\beta_2 \rightarrow \infty$ then both the coefficient of the nonlinear derivative terms in Eq.~(\ref{poisson+eom-2}) and $r_V$ vanish. In this case, the additional gravitational force is not suppressed below any distance scale. This shows how the derivative interactions of the scalar field are able to suppress the fifth force.

The lower panel of Fig.~\ref{beta-rv} shows the time evolution of the Vainshtein radius, $r_V$, for the Cubic Galileon and DGP models. In the latter we have assumed that $r_c = H_0^{-1}$, where $r_c$ is the DGP crossover scale\footnote{Very briefly, the crossover scale, $r_c$, is a parameter of the DGP model that gives a measure of the length scale below which gravity is four-dimensional and above which it is five-dimensional \cite{2000PhLB..485..208D}.}, so that $r_V$ could be plotted for the two models with the same units of $r_SH_0^{-2}$. At early times, $r_V$ is very small which means that there is no screening, or that only very small scales are screened. As the universe expands, the range of the screening increases but it does so at different rates in the two models shown. In particular, at $a \approx 0.5$, the Vainshtein radius of a given matter source in the Cubic Galileon model becomes comparable to that in the DGP model. From $a \approx 0.5$ until the present time, the Vainshtein radius in the Cubic Galileon model is larger than in the DGP model, with the values differing by approximately one order of magnitude at $a = 1$. In practice, this means that in the Cubic Galileon model, the fifth force resulting from a given matter source at $a = 1$ is screened out to a distance which can be about ten times larger than in the DGP model.

\subsection{Model parameters}

\begin{table}
\caption{Parameters of the Cubic Galileon model that best fits the data from the WMAP 9yr results \cite{Hinshaw:2012fq}, the SNLS 3yr sample \cite{Guy:2010bc} and the BAO measurements from the 6dF Galaxy Survey \cite{Beutler:2011hx}, from the SDSS DR7 \cite{Percival:2009xn} and from the SDSS-III BOSS \cite{Anderson:2012sa}. $\Omega_{r0}$, $\Omega_{b0}$, $\Omega_{c0}$, $h$, $n_s$, and $\tau$ are, respectively, the present day fractional energy density of radiation ($r$), baryons ($b$) and cold dark matter ($c$), the dimensionless present day Hubble expansion rate, the primordial spectral index and the optical depth to reionization. The scalar amplitude at recombination $A_s$ refers to a pivot scale $k = 0.02 \rm{Mpc}^{-1}$. In this model we assume that the universe is spatially flat. The parameters $c_2$, $c_3$ are the dimensionless constants in the Cubic Galileon Lagrangian (Eq.~\ref{Galileon action}) and $\rho_{\varphi, i} / \rho_{m,i}$ is the ratio of the Galileon and total matter ($m$) energy densities at $z_i$. We also show the value of $\chi^2 = -2{\rm log}P$ (where $P$ is the posterior probability obtained from the data), the Galileon field time derivative at $z_i$, the age of the Universe and the value of $\sigma_8$ at $z = 0$. Only in this table, the subscript "$_i$" refers to quantities evaluated at $z = z_i = 10^6$. }
\begin{tabular}{@{}lccccccccccc}
\hline\hline
\\
Parameter  & \ \ WMAP9+SNLS+BAO & \ \ 
\\
\hline
\\
$\chi^2$                                                               			         &\ \ $8006.50$ & \ \ 
\\
$\Omega_{r0}{h}^2$                                				&\ \ $4.28\times10^{-5}$ & \ \ 
\\
$\Omega_{b0}{h}^2$                                				&\ \ $0.02196$ & \ \ 
\\
$\Omega_{c0}{h}^2$                                 				&\ \   $0.1274$ & \ \ 
\\
${h}$                                                                			&\ \   $0.7307$ & \ \ 
\\
$n_s$                                                                 			           &\ \    $0.953$ & \ \ 
\\
$\tau$                                                                     			&\ \   $0.0763$ & \ \
\\
$\rm{log}\left[ 10^{10}A_s \right]$    			&\ \  $3.154$ & \ \ 
\\
$\rm{log}\left[\rho_{\varphi, i} / \rho_{m,i}\right]$         &\ \   $-4.22$ & \ \ 
\\
$c_2 / c_3^{2/3}$                                                                                   &\ \    $-5.378$ & \ \
\\
$c_3$                                                                                   &\ \    $10$ & \ \
\\
\hline
\\
$\dot{\bar{\varphi}}_i c_3^{1/3}$					  		&\ \  $1.104\times10^{-9}$ & \ \
\\
Age (Gyr)					  		&\ \ $13.748$ & \ \
\\
$\sigma_8(z = 0)$					  		&\ \ $0.997$ & \ \
\\
\hline
\hline
\end{tabular}
\label{table-max}
\end{table}

\begin{figure}
	\centering
	\includegraphics[scale=0.42]{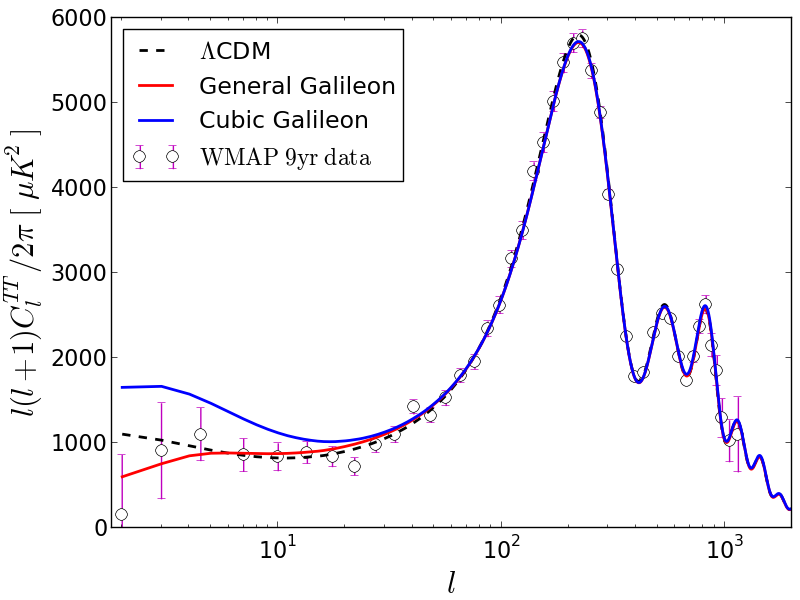}
	\caption{Cosmic microwave background temperature angular power spectrum, plotted as function of the multipole moment $l = \pi/\theta$ (where $\theta$ is the angle on the sky), for the Cubic Galileon model of Table \ref{table-max} (solid blue) together with the WMAP 9yr data (open circles with errorbars). The power spectra of the best-fitting general Galileon gravity model to WMAP9+SNLS+BAO data \cite{Barreira:2013jma} (solid red) and the $\Lambda$CDM model of \cite{Hinshaw:2012fq} (dashed black) are also shown for comparison. At $l = 2$, from top to bottom, the lines correspond, respectively, to the Cubic Galileon, $\Lambda$CDM and general Galileon model.}
\label{Cl}\end{figure}

The Cubic Galileon model we simulate in this paper is the best-fitting model to a combination of data comprising the WMAP 9yr results \cite{Hinshaw:2012fq}, the Supernovae Legacy Survey 3yr sample \cite{Guy:2010bc} and the BAO measurements from the 6df Galaxy Survey \cite{Beutler:2011hx}, the Sloan Digital Sky Survey (SDSS) DR7 \cite{Percival:2009xn} and the SDSS-III Baryon Oscillation Spectroscopic Survey (BOSS) \cite{Anderson:2012sa}\footnote{From here on we shall refer to this combined dataset as WMAP9+SNLS+BAO.}. The model parameters, shown in Table \ref{table-max}, were obtained by following the strategy presented in \cite{Barreira:2013jma}, in which we placed observational constraints on the most general Covariant Galileon gravity model by running Monte Carlo Markov chains, and to which we refer the reader for further details. In Fig.~\ref{Cl}, we show the CMB temperature power spectra of the Cubic Galileon (solid blue) and general Galileon (solid red) models that best fit the WMAP9+SNLS+BAO data. We also show the power spectrum of the WMAP 9yr \cite{Hinshaw:2012fq} best-fitting $\Lambda$CDM model for comparison. The top panel of Fig.~\ref{H-G-delta} shows the Hubble expansion rate of the best-fitting Cubic Galileon model.

In \cite{Barreira:2013jma} we showed that the general Covariant Galileon model can provide a fit to the data which is of comparable quality to that of the  $\Lambda$CDM model. In fact, if only the WMAP9 data is considered in the constraints, then the Galileon model performs better than the standard $\Lambda$CDM. However, in Fig.~\ref{Cl}, we see that the Cubic Galileon model is not able to provide as good a fit at low $l$ as the more general Galileon theory. In the Cubic Galileon model there is substantially more power on the largest angular scales, where the integrated Sachs-Wolfe (ISW) effect plays the dominant role in determining the shape of the spectrum. This seems to indicate that the success of the general Galileon model in fitting the observational data may lie in the detailed interplay of all the different Galileon Lagrangian densities. In other words, if one does not consider all of the Lagrangian terms, then the model will lose some of its ability to fit the data. In particular, the difference in $\chi^2 = -2{\log} P$ (where $P$ is the posterior probability) between the best-fitting Cubic and the general Galileon models is $\chi^2_{\rm{Cubic}} - \chi^2_{\rm{general}} \approx 16.5$, for the WMAP9+SNLS+BAO dataset. This fact alone may already be sufficient to rule out the Cubic Galileon gravity model. A detailed study of the way the different Lagrangians of the general Galileon model "interact" with one another to provide a reasonable fit to the data is beyond the scope of the present paper and will be left for future work.

In \cite{Barreira:2012kk, Barreira:2013jma}, we showed that, at the linear level in perturbation theory and hence without considering any screening effects, the general Galileon model typically predicts too much matter clustering to be compatible with the current data. However, we pointed out that such results were subject to the validity of linear theory on the relevant scales. We further argued that a better understanding of the nonlinearities in the Galileon model was necessary before making definitive statements about the observational merit of the theory. In the following sections, as an initial step to build intuition about the more general Galileon model, we study the case of the simpler Cubic Galileon model despite its somewhat poorer performance in fitting the WMAP9+SNLS+BAO data, compared with the general Galileon model. The study of the nonlinearities in the general Galileon theory will be left for future work.

\subsection{Linear growth of the density field}\label{linear-section}

\begin{figure}
	\centering
	\includegraphics[scale=0.49]{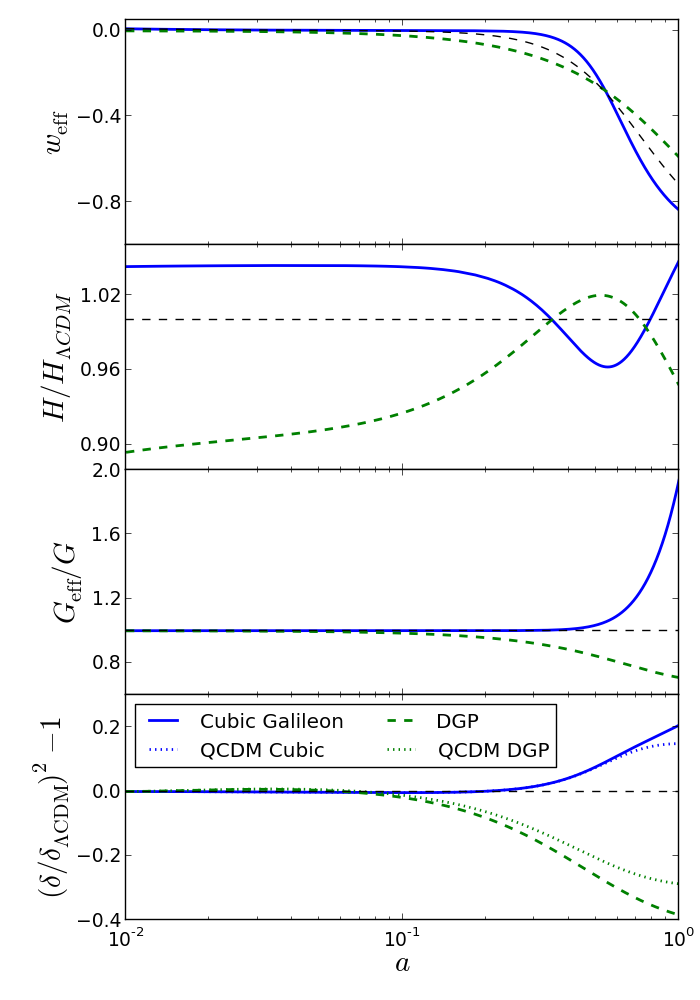}
	\caption{From top to bottom, the first three panels show, respectively, the time evolution of the effective cosmological equation of state parameter $w_{\rm eff} = -1 - 2\dot{H}/(3H^2)$, the ratio of the Hubble expansion rate of the Cubic Galileon (solid blue) and DGP (dashed green) models to the $\Lambda$CDM model (dashed black), $H/H_{\Lambda\rm{CDM}}$, and the time evolution of the effective gravitational constant $G_{\rm{eff}}$. The bottom panel shows the time evolution of the relative difference of the square of the density contrast in the Cubic Galileon (solid blue), $\rm{QCDM}_{\rm{Cubic}}$ (dotted blue), DGP (dashed green) and $\rm{QCDM}_{\rm{DGP}}$ (dotted green) models to the $\Lambda$CDM model, $(\delta/\delta_{\Lambda\rm{CDM}})^2 - 1$. The Cubic Galileon model plotted is the model of Table \ref{table-max}, the DGP model is the model simulated in \cite{Li:2013nua} and the $\Lambda$CDM model is the best-fitting model from \cite{Hinshaw:2012fq}. In the bottom panel, at $a = 1.0$, from top to bottom, the lines correspond to the Cubic Galileon, $\rm{QCDM}_{\rm{Cubic}}$, $\rm{QCDM}_{\rm{DGP}}$ and DGP model, respectively.}
\label{H-G-delta}\end{figure}

Since one of our goals is to test the validity of linear perturbation theory on sub-horizon scales in the Cubic Galileon model, it is instructive to first study the evolution of the density fluctuations in linear perturbation theory. The time evolution of a small sub-horizon matter density fluctuation characterized by the density contrast, $\delta = \delta\rho_m/\bar{\rho}_m = \left(\rho_m - \bar{\rho}_m\right)/\bar{\rho}_m$, is governed by

\bq
\label{delta-equation}\delta'' + \left[3 + a\frac{H'}{H}\right]\frac{\delta'}{a} - \frac{3\Omega_{m0}H_0^2}{2H^2}\frac{G_{\rm{eff}}}{G} \frac{\delta}{a^5}= 0,
\eq
where a prime denotes a derivative with respect to the cosmic scale factor, $a$. In the linear regime, each mode of the perturbed density field evolves independently and the evolution is determined by the expansion rate, $H$, the matter density, $\Omega_{m0}h^2$, and $G_{\rm{eff}}$. $H$ is obtained by solving Eqs.~(\ref{Friedman1}), (\ref{Friedman2}) and (\ref{background-EoM}), whereas $G_{\rm{eff}}$ is the time-dependent effective gravitational constant, which can be obtained from the linearized Poisson equation ($\beta_1 \rightarrow \infty,\ \beta_2\ \rm{finite}$):

\bq
\label{linear-poisson}\partial^2\Phi = 4\pi G_{\rm{eff}}a^2\delta\rho_m,
\eq
where
\bq
\label{G_eff}G_{\rm{eff}} = G\left(1 - \frac{2}{3}\frac{c_3\dot{\varphi}^2}{M_{\rm{Pl}}\mathcal{M}^3\beta_2}\right).
\eq

In Fig.~\ref{H-G-delta}, from the top to the bottom panels, we plot the time evolution of $w_{\rm eff} = -1 - 2\dot{H}/(3H^2)$, $H$, $G_{\rm{eff}}$ and the relative difference of the square of the linear density contrast compared to that in $\Lambda$CDM, $(\delta/\delta_{\Lambda\rm{CDM}})^2 - 1$, for the Cubic Galileon (solid blue) and DGP (dashed green) models. The $\Lambda$CDM model used here is the best-fitting model to the WMAP 9yr data \cite{Hinshaw:2012fq}. To solve Eq.~(\ref{delta-equation}) we use, for all models, the initial conditions $a_i = 0.01$, $\delta(a = a_i) = a_i$ and $\delta'(a = a_i) = 1$, which correspond to the matter dominated era solution, $\delta \propto a$. We also solve the evolution of $\delta$ for two other models which we call $\rm{QCDM}_{\rm{Cubic}}$ and $\rm{QCDM}_{\rm{DGP}}$. In these models $G_{\rm{eff}} = G$, but $H$ and $\Omega_{m0}h^2$ are taken to be those of the Cubic Galileon and DGP models, respectively. This allows us to disentangle the relative importance of the modified gravitational strength in changing the growth of the linear matter density field.

At early times, $a \lesssim 0.1$, $G_{\rm{eff}} \approx 1$ in both models. Therefore, any differences from $\Lambda$CDM are driven by the modified expansion history, $H$, and different matter densities $\Omega_{m0}h^2$. The former acts as a frictional force that slows down linear structure growth, while the latter has the opposite effect. In the Cubic Galileon model, both $H$ and $\Omega_{m0}h^2$ are larger than in $\Lambda$CDM, and their effects cancel out. The same thing happens in the DGP model, but because both $H$ and $\Omega_{m0}h^2$ are smaller. As a result, for $a \lesssim 0.1$, the relative difference in the evolution of $(\delta/\delta_{\Lambda\rm{CDM}})^2 - 1$ is almost zero for both models.

During the epochs when $a \gtrsim 0.1$, the evolution of $\delta$ is determined by the interplay of the modifications in $H$, $\Omega_{m0}h^2$ and $G_{\rm{eff}}$. In the lower panel of Fig.~\ref{H-G-delta}, we see that the modifications to gravity enhance structure formation at late times in the Cubic Galileon model, but suppress it in the DGP model. Also, the amount by which the growth is enhanced in the Cubic Galileon model is smaller than the amount by which it is suppressed in the DGP model. In both models, the dominant role in modifying the growth of the linear density field relative to $\Lambda$CDM is played by the modified matter density. At $a = 1$, the difference relative to $\Lambda$CDM of the squared density contrast is, approximately, $15\%$ and $-28\%$ in $\rm{QCDM}_{\rm{Cubic}}$ and $\rm{QCDM}_{\rm{DGP}}$, respectively. Taking into account the modified gravitational strength in the Cubic Galileon model adds only about 5$\%$ to the relative difference to $\Lambda$CDM at $a = 1$. This happens despite the fast growth of $G_{\rm{eff}}$ in the Cubic Galileon model after $a \approx 0.6$, with its value being almost twice as large as the corresponding value in GR at the present day. In the case of the DGP model, the effect of the gravitational strength is more pronounced and it accounts for about 10$\%$ of the difference from $\Lambda$CDM at $a = 1$. The reason here is that the value of $G_{\rm{eff}}/G$ starts to deviate from unity at much later times in the Cubic Galileon model ($a \gtrsim 0.5$) than in the DGP model ($a \gtrsim 0.1$). As a result, in the Cubic Galileon model, there is less time for the modified $G_{\rm{eff}}$ to change the linear growth of structure by as much as is seen in the DGP model. These later-time modifications of the gravitational force in the Cubic Galileon model constitute an important difference relative to other modified gravity models that have been simulated recently \cite{Li:2013nua, Wyman:2013jaa}. As we will see next, this feature plays a decisive role in determining the efficiency of the screening mechanism.

\section{N-body simulations}\label{simulations-section}

\subsection{N-body equations}

The N-body codes {\tt RAMSES} \cite{Teyssier:2001cp} and {\tt ECOSMOG} \cite{Li:2011vk} use "super-comoving" coordinates \cite{Martel:1997hk}:

\bq
d\tilde{x} &=& \frac{dx}{aL};\ \ \ \ \ \ \ \ \ \ \ \ \ \ \ \   d\tilde{t} = \frac{H_0dt}{a^2}; \nonumber \\
\ \ \ \tilde{v} &=& \frac{av}{LH_0};\ \ \ \ \ \ \ \ \ \ \ \ \ \ \tilde{\Phi} = \frac{a^2\Phi}{\left(LH_0\right)^2}; \nonumber \\
\ \ \tilde{\rho} &=& \frac{\rho a^3}{\rho_{c0}\Omega_{m0}}; \ \ \ \ \ \ \ \ \ \ \tilde{\varphi} = \frac{a^2\varphi}{\left(LH_0\right)^2M_{\rm{Pl}}},
\eq
(quantities with tildes are those in the super-comoving system), where $dx$ and $dt$ are the differential interval of physical space and time, $\rho_{c0} = 8\pi G/(3H_0^2)$ is the critical density today and $\Omega_{m0} = \Omega_{b0} + \Omega_{c0}$ is the density of matter (baryons and cold dark matter) in units of $\rho_{c0}$. $v$ is the particle velocity and $L$ is the size of the simulation box in units of $\rm{Mpc} / \it{h}$. In this coordinate system the background matter density is unity, $\tilde{\bar{\rho}} = 1$. Note that under such a coordinate transformation all quantities become dimensionless.

In super-comoving coordinates, Eqs.~(\ref{poisson-2}) and ({\ref{poisson+eom-2}}) are given, respectively, by:

\bq
\label{poisson-code}\tilde{\partial}^2\tilde{\Phi} = \frac{3}{2}\Omega_{m0}a\left(\tilde{\rho} - 1\right) - \frac{\kappa c_3}{\mathcal{M}^3}\frac{\beta}{\beta_2}\dot{\varphi}^2\tilde{\partial}^2\tilde{\varphi}, \\
\label{poisson+eom-code}\tilde{\partial}^2\tilde{\varphi} + \frac{1}{3\left(\beta_1\beta_2/\beta\right)a^4}\left[(\tilde{\partial}^2{\tilde{\varphi}})^2 - (\tilde{\partial}_i\tilde{\partial}_ j{\tilde{\varphi}})^2\right]  \nonumber \\
= \frac{\Omega_{m0}a}{\beta}\left(\tilde{\rho} -1 \right).
\eq
As in the derivation of Sec.~\ref{vainshtein}, Eq.~(\ref{poisson+eom-code}) can be viewed as a second-order algebraic equation for $\partial^2\varphi$ (from here on we will omit the tildes for ease of notation). To avoid numerical problems related to the choice of which branch of solutions to take, we first solve this equation analytically to obtain:

\bq
\label{algebraic-solution-3}\partial^2\varphi = \frac{-\alpha \pm \sqrt{\alpha^2 + 4\left(1-\varepsilon\right)\Sigma}}{2\left(1-\varepsilon\right)},
\eq
in which

\bq
\label{alpha}\alpha &\equiv& 3\left(\beta_1\beta_2/\beta\right)a^6, \\
\label{Sigma}\Sigma &\equiv& ({\partial}_i{\partial}_ j{{\varphi}})^2 + \frac{\alpha\Omega_{m0}a^3}{\beta}\left(\rho - 1\right) - \varepsilon\left(\partial^2\varphi\right),
\eq
and we have followed the strategy of \cite{Chan:2009ew, Li:2013nua}, to use the operator-splitting trick to avoid problems associated with imaginary square roots and where $\varepsilon$ is a free constant coefficient we choose to be $1/3$\footnote{The solution is obtained by solving the equivalent equation $(1-\varepsilon)\left(\partial^2\varphi\right)^2 + \alpha\left(\partial^2\varphi\right) - \Sigma = 0$.}. The choice of the solution branch is determined by the condition that the physical result that $\partial^2\varphi \rightarrow 0$, when $\rho \rightarrow 1$, should be recovered, i.e., if there are no density fluctuations then there should be no fifth force \cite{Li:2013nua}. As a result, one should choose the sign of the square root in Eq.~(\ref{algebraic-solution-3}) to be the sign of $\alpha$, or equivalently, the sign of $\beta_1\beta_2/\beta$. With such a choice, Eq.~(\ref{algebraic-solution-3}) can be written as

\bq
\label{algebraic-solution-4}\partial^2\varphi = \frac{-\alpha + \rm{sign}(\alpha) \sqrt{\alpha^2 + 4\left(1-\varepsilon\right)\Sigma}}{2\left(1-\varepsilon\right)}.
\eq
To determine the particle trajectories, the N-body code first solves the Galileon field equation (Eq.~(\ref{algebraic-solution-4})) to determine $\partial^2\varphi$. The solution is then plugged into the Poisson equation (Eq.~(\ref{poisson-code})), which is solved to obtain the gradient of $\Phi$, which gives the total force (GR + fifth force) under which the simulation particles move.

The discretization of Eqs.~(\ref{poisson-code}) and (\ref{algebraic-solution-4}) is identical to the case of the DGP model (apart from the different coefficients, c.f. Sec.\ref{quasi-static-section}). Such equations are lengthy and were already presented in \cite{Li:2013nua}, to which we refer the interested reader.

\subsubsection{Problems with imaginary square roots}

The quadratic nature of Eq.~(\ref{poisson+eom-code}) raises the possibility that, under some circumstances, there might not be real solutions. If we look again at the case of spherical symmetry, from Eqs.~(\ref{algebraic-solution-1}), (\ref{algebraic-solution-2}) and (\ref{Vainshtein-radius}) we see that the condition for the existence of real solutions is given by:

\bq
\label{discriminant} \Delta \equiv 1 + \frac{1}{\beta_1\beta_2}\frac{64\pi G M_{\rm{Pl}}}{9\mathcal{M}^3r^3}\int_0^r\delta\rho_{m}(\xi)\xi^2d\xi \ge 0.
\eq
This equation shows that in low density regions, such as voids, where $\delta\rho_{m} < 0$, it is possible for $\Delta$ to be negative (note that $\beta_1\beta_2 > 0$). In fact, this is exactly what we have found in our simulations of the Cubic Galileon model: for $a \gtrsim 0.8$, there are regions in the simulation boxes for which there are no real solutions for the fifth force. Such a problem, nevertheless, is absent from the DGP simulations performed with the same N-body code \cite{Li:2013nua}. The reason is primarily related to the different time evolution of the quantity $\beta_1\beta_2/\beta$ (or equivalently $\beta_1\beta_2$) in both models. Looking at Eq.~(\ref{discriminant}), one sees that the smaller the value of $\beta_1\beta_2$, the easier it is for $\Delta$ to be negative in low density regions. In Fig.~\ref{beta-rv}, we have seen that at late times, $\beta_1\beta_2$ is smaller in the Cubic Galileon model than it is in the DGP model, which is why the imaginary square root problem shows up in the former and not in the latter.

This problem is likely to be a consequence of the quasi-static limit approximation. The terms we have neglected while deriving the quasi-static field equations in Sec.~\ref{equations} may not be completely negligible in certain circumstances, such as when the matter density is low. In particular, such terms might be the missing contribution to Eq.~(\ref{algebraic-solution-4}) that prevents the imaginary solutions. In the simulations for this paper, we work our way around this problem by simply setting $\Delta = 0$ whenever this quantity becomes negative. Such a solution, although crude and not theoretically self-consistent, should not have a measurable impact on the small scale nonlinear matter power spectrum and halo mass functions. The reason is that the clustering power on small scales is dominated by the high density regions where the problem does not appear. 

We stress however that even if the fifth force would not have become imaginary, one would still have an inaccurate calculation in low densities because the quasi-static limit is not expected to be a good approximation there. For instance, this is the case of the DGP simulations that have been performed so far \cite{Schmidt:2009sg, Schmidt:2009sv, Chan:2009ew, Schmidt:2009yj, Khoury:2009tk, Li:2013nua}. The case of the Cubic Galileon is more severe because it forces one to fix some terms in the equations in an {\it ad hoc} way. It should be noted that it is not clear that our solution to the imaginary fifth force problem is making the calculation more inaccurate than in the DGP simulations. To fully address this question one would have to simulate the full model equations (i.e.~without assuming the quasi-static limit), which is beyond the scope of the present paper.

\subsection{Outline of the code algorithm}

In {\tt ECOSMOG}, the equations are solved on a grid that adaptively refines itself in regions where the particle density is high. The starting point is a fixed regular mesh that fills the entire simulation domain (called the domain grid). This grid gets refined in regions where the number of matter particles per cell exceeds a given threshold value, $N_{p,th}$. The refinement process only stops after all the cells in the refinements contain a number of particles which is smaller than $N_{p, th}$. Each refinement level is used to compute the force on the particles that are within its domain, with the boundary conditions being set via interpolation from the coarser grid levels. A good feature of AMR codes in cosmological simulations is that, in this way, one can attain sufficiently high resolution in high matter density regions, while saving computational effort in low-density regions where the resolution can be lower.

The {\tt ECOSMOG} code differs from {\tt RAMSES} by having a solver for the equations of the extra scalar degree of freedom that performs Gauss-Seidel iterative relaxations on the grid. The algorithm also employs the multigrid technique to overcome the slow convergence of the relaxations after a few iterations. In brief, at a given refinement level, the Gauss-Seidel relaxations are efficient at damping only the Fourier modes of the error that are comparable to the cell size of that level. This leads to poor convergence on that level since the longer-wavelength modes are barely reduced. Using the multigrid technique, the code interpolates the equations to the next coarser level on which they are solved. This allows for the longer wavelength modes of the error to decay much more quickly. The coarser solution is then interpolated back to the finer initial level to correct the solution.

The code features that are most relevant for modified gravity simulations are explained in detail in \cite{Li:2011vk}.

\subsection{Simulation details}

\begin{table}
\caption{Summary of the three variants of the Cubic Galileon model we simulate in this paper. All models have the same background expansion history and only differ in the force law.}
\begin{tabular}{@{}lccccccccccc}
\hline\hline
\\
Model  & \ \ $H(a)$ & Total Force &\ \ 
\\
\hline
\\
Full Cubic Galileon                    &\ \  Cubic Galileon &  \ \ GR + Screened fifth force &\ \ 
\\
Linear Cubic Galileon               &\ \ Cubic Galileon &  GR + linear fifth force &\ \ 
\\
$\rm{QCDM}_{\rm{Cubic}}$   &\ \ Cubic Galileon &  GR&\ \ 
\\
\hline
\hline
\end{tabular}
\label{table-models}
\end{table}

We have simulated three variants of the Cubic Galileon model, summarized in Table \ref{table-models}. The first one is the "full" Cubic Galileon model characterized by Eqs.~(\ref{Friedman1}), (\ref{Friedman2}), (\ref{background-EoM}), (\ref{poisson-2}) and (\ref{poisson+eom-2}). The second one is a linear force model, in which, instead of solving Eqs.~(\ref{poisson-2}) and (\ref{poisson+eom-2}), we use the linearized Poisson equation, Eq.~(\ref{linear-poisson}). Simulations of these two models allow us to measure the effectiveness of the Vainshtein mechanism in suppressing the fifth force. Finally, the third model is the $\rm{QCDM}_{\rm{Cubic}}$ model of Fig.~\ref{H-G-delta} (bottom panel, dotted blue line), where there is no fifth force but the background expansion history is that of the Cubic Galileon model. As in Sec.~\ref{linear-section}, the simulations of this model allow one to isolate the changes introduced by the modified force law, and exclude those arising from the modified expansion history. The effects of the latter are small on nonlinear scales and its effects on large scales, which evolve linearly, have already been shown in Fig.~\ref{H-G-delta}.

The initial conditions of the simulations were generated using the {\tt MPGRAFIC} software \cite{Prunet:2008fv} at $a = 0.02$ ($z = 49$). To generate the initial conditions we use the matter power spectrum at $a = 0.02$ of the Cubic Galileon model of Table \ref{table-max}\footnote{In the generation of the initial conditions with {\tt MPGRAFIC}, we have assumed a flat $\Lambda$CDM model with the same $\Omega_{m0}$ as in Table \ref{table-max}. Using the {\tt CAMB} code, we have explicitly checked that the linear matter power spectrum of this $\Lambda$CDM and Cubic Galileon models is the same at the sub-percent level.}. All three models use the same initial conditions and thus, the differences between them are purely due to the differences in the force calculation and not due to any mismatch between the phases of the initial density field. We use two box sizes $L = 200\rm{Mpc}/\it{h}$ and $L = 400\rm{Mpc}/\it{h}$, with a refinement criteria of $N_{p, th} = 8$ and $N_{p, th} = 6$, respectively. For both box sizes the total number of particles is $N_p = 512^3$, and the domain grid has $2^9 = 512$ cells in each direction. The finest refinement level, if it were to cover the whole simulation volume, would have in each direction, $2^{15}$ and $2^{16}$ cells for the larger and smaller boxes, respectively. To allow for statistical averaging and overcome the sample variance in the generation of the initial conditions, for each box we generate five realizations of the same initial density field by using different random seeds. We have therefore run 30 simulations in total: 5 per cosmological model per box size. For all these simulations, the convergence criterion of the Gauss-Seidel algorithm was the same as that used in \cite{Li:2013nua}.

\section{Results}\label{results-section}

\subsection{Matter power spectrum results}\label{pk-section}

\begin{figure*}
	\centering
	\includegraphics[scale=0.39]{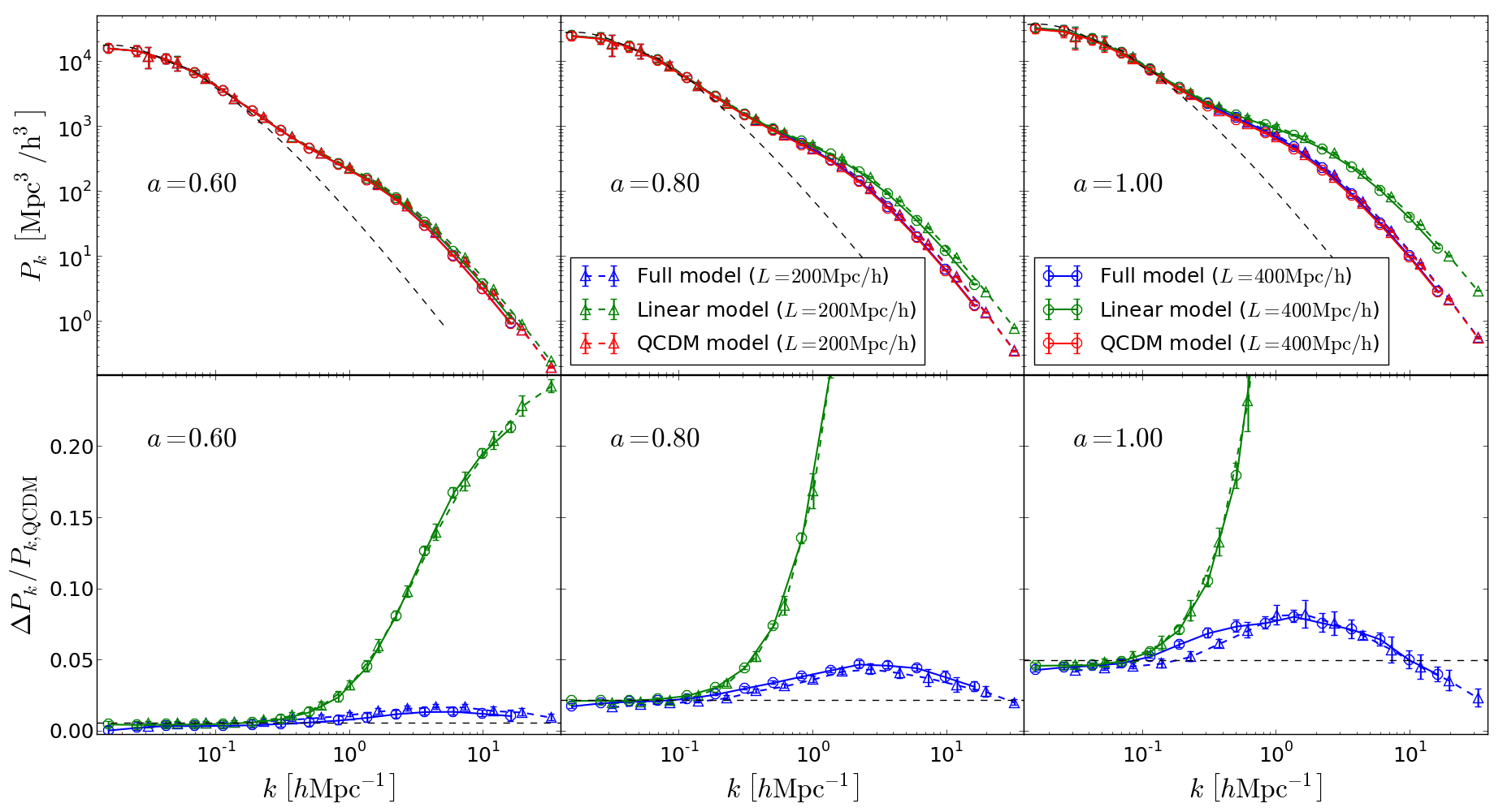}
	\caption{The top panels show the simulation matter power spectrum $P_k$ of the $L = 200\rm{Mpc}/\it{h}$ (dashed lines and open triangles) and $L = 400\rm{Mpc}/\it{h}$ (solid lines and open circles) boxes for three output times ($a = 0.6$, $a = 0.8$ and $a = 1.0$, from left to right) for the Cubic Galileon (blue),  linear-force Cubic Galileon (green) and $\rm{QCDM}_{\rm{Cubic}}$ (red) models. The bottom panels show the relative difference of the matter power spectrum of the Cubic Galileon and linear-force Galileon to the $\rm{QCDM}_{\rm{Cubic}}$ models, $\Delta P_k/P_{k, {\rm QCDM}} \equiv  (P_k - P_{k, {\rm QCDM}})/P_{k, {\rm QCDM}}$. The dotted black lines correspond to the linear perturbation theory result and were obtained with a modified version of the {\tt CAMB} code \cite{camb_notes, Barreira:2012kk}. The results shown are obtained by averaging over the simulations of the different initial conditions realizations and the errorbars show the standard deviation over these realizations.}
\label{nonlinear-pk}\end{figure*}

In the top panels of Fig.~\ref{nonlinear-pk}, we show the matter power spectra, $P_k = \langle |\delta_k|^2 \rangle$, measured from the simulations of the Cubic Galileon model (blue) and of the corresponding linear-force (green) and $\rm{QCDM}$ (red) models, for the $L = 200\rm{Mpc}/\it{h}$ (dashed lines and open triangles) and $L = 400\rm{Mpc}/\it{h}$ (solid lines and open circles) boxes. The bottom panels show the relative difference of the Cubic Galileon and linear-force models to the QCDM model, $\Delta P_k/P_{k, \rm{QCDM}} \equiv  (P_k - P_{k, \rm{QCDM}})/P_{k, \rm{QCDM}}$. From left to right, the panels correspond to the results at three different epochs: $a = 0.6$ ($z = 0.67$), $a = 0.8$ ($z = 0.25$) and $a = 1$ ($z = 0$). The nonlinear matter power spectrum is measured using the {\tt POWMES} code \cite{Colombi:2008dw}.

On the largest scales probed by the simulations, $k \lesssim 0.1 h\  \rm{Mpc}^{-1}$, we find that there is good agreement between the nonlinear and the linear theory results. On these scales, the power spectra of the Cubic Galileon and the linear-force model agree to within $\sim 5\%$ with linear perturbation theory (dotted lines). Such an agreement leads to two conclusions. The first is that it shows that the quasi-static approximation we applied in deriving the simulation equations in Sec.~\ref{equations} does not have an impact on large scales, where $\|\delta\|\ll 1$. Otherwise, one would expect the linear-force model (green lines) to disagree with the linear theory calculation, which does not happen. The second conclusion relates to the range of validity of linear theory in the Cubic Galileon model. As we said above, one of the goals of this paper is to test the degree to which the inclusion of the screening effects modifies the linear theory predictions. Figure \ref{nonlinear-pk} shows that the power spectrum of the Cubic Galileon model recovers the linear theory expectation on scales $k \lesssim 0.1 h\ \rm{Mpc}^{-1}$, for all of the output times shown. That is, in the Cubic Galileon model, the screening mechanism does not affect the large scales typically associated with linear perturbation theory, which is therefore still a valid approximation.

On smaller scales ($k \gtrsim 0.1 h \rm{Mpc}^{-1}$), the different Fourier modes of the density fluctuations no longer evolve independently and linear theory ceases to be valid. It is well known that the nonlinearities in the density field result in an enhancement of structure formation as can be seen in the top panel of Fig.~\ref{nonlinear-pk}. Such an enhancement is much more pronounced in the linear-force model (green lines) due to the large gravitational strength (c.f. Fig.~\ref{H-G-delta}). In Sec.~\ref{linear-section}, we saw that the modified $G_{\rm{eff}}$ had a modest influence over enhancing the growth of the linear density field due to the deviations from the standard value $G$ only appearing at late times. However, on small scales, the effects of the unscreened fifth force are highly amplified by the strong mode-coupling of the density fluctuations on different scales. For instance, at $k = 10h \rm{Mpc}^{-1}$, the relative difference between the linear-force model and QCDM is $\sim 100\%$ at $a = 0.8$ and $\sim 300\%$ at $a = 1.0$ (not shown in the scales of the bottom middle and bottom right panels). As a double check of the strong enhancement of the small scale clustering power in the linear force model, we have re-simulated one of the initial condition realizations using the {\tt Gadget}-2 TreePM code \cite{Springel:2005mi} \footnote{This was done by interpolating the values of $H$ and $G_{\rm eff}/G$ from a table in {\tt Gadget}.}. The results of the latter agree very well with those from {\tt ECOSMOG}.

Such an enhancement in the amplitude of clustering is very hard to reconcile with the current observations, a conclusion which highlights the vital role that screening mechanisms play in modified gravity theories. Here, in the particular case of the Cubic Galileon model, we see that the screening mechanism works very effectively to restrain the fifth force from boosting the formation of structure by a significant factor. For instance, at $a = 1$ ($a = 0.8$), the screened fifth force (blue lines) is only responsible for a relative increase which is less than $\sim 8\%$ ($\sim 5\%$) on all scales. This efficiency of the screening is tied to the fact that the fifth force only starts to become important at relatively late times (c.f. Fig.~\ref{H-G-delta}), when the Vainshtein radius is sufficiently large (c.f.~Fig.~\ref{beta-rv}) for the fifth force to be screened on scales $k \gtrsim 0.1 h\rm{Mpc}^{-1}$.

At $a = 1$, on scales $0.1 h\rm{Mpc}^{-1} \lesssim \it{k} \lesssim \rm{0.8} \it{h}\rm{Mpc}^{-1}$, we see that the relative differences in the power of the full model predicted using the two boxes (bottom right panel, blue lines) do not show the same level of agreement seen at other epochs or scales. In particular, the fifth force is better screened in the smaller box ($L = 200\rm{Mpc}/\it{h}$), and a possible explanation lies in resolution effects. On these mildly nonlinear scales, the density field is smoother than on smaller scales, and therefore, the density peaks are not as pronounced. The smaller box has a higher mass resolution, and consequently, is expected to resolve better the lower density peaks on these scales. The effects of the screening are then more accurately captured in the smaller box, which results in a slightly lower clustering amplitude. \footnote{On smaller scales, the density peaks are more pronounced, and therefore, easier to resolve. This explains why on smaller scales the two boxes agree much better.}. Although less significantly, at $a = 0.8$, there is also the same trend for the smaller box to have a better screened fifth force. 

In the bottom panels of Fig.~\ref{nonlinear-pk}, we see that the relative difference of the Cubic Galileon model to $\rm{QCDM}_{\rm{Cubic}}$ increases from large to small scales until it reaches a peak, decreasing thereafter. This is because the fifth force is screened on smaller scales, and therefore, can only increase the power on larger scales \footnote{On scales larger than the peak, from large to small scales, the relative difference increases due to the fact that the effects of the (unscreened) fifth force are more pronounced on smaller scales.}. In particular, the smallest scales probed by the simulations, $k \gtrsim 20 h\ \rm{Mpc}^{-1}$, have always been screened, and as a result, the relative difference barely evolves with time. On these small scales, standard gravity is always recovered. In addition, in Fig.~\ref{nonlinear-pk} one also sees that the position of the peak of the relative difference moves from smaller to larger scales with time. The reason now is related to a combination of two factors. The first one is that, over time, the nonlinear mode-coupling effects become progressively more important at larger scales, which therefore will cluster faster. The second factor relates to the time evolution of the Vainshtein radius, which becomes larger with time (c.f.~Fig.~\ref{beta-rv}). As a result, scales that were previously affected by the fifth force will eventually become screened. These two factors together move the peak from smaller to larger scales.

Recently, \cite{Li:2013nua, Wyman:2013jaa} have performed high-resolution N-body simulations of two models similar to the Cubic Galileon model. In these works, it was found that the effects of the fifth force were more pronounced than in the case presented here. In particular, in \cite{Wyman:2013jaa} it was found that the screened fifth force leads to an increase of the clustering on small scales ($k \sim 1~h\rm{Mpc}^{-1}$) that can be as large as $50\%$ by the present time (c.f. Fig. 4 therein). A similar result is found in the case of the DGP model \cite{Li:2013nua}, where the screened fifth force can decrease the clustering by $\sim 18\%$ at the present time (c.f. Fig. 3 therein). The reason why the screening mechanism is not as effective in these models as it is in the Cubic Galileon model is once again related to the fact that, in the latter, the fifth force becomes important only at late times. In the case of the models simulated in \cite{Li:2013nua} and \cite{Wyman:2013jaa}, the linear fifth force represents a sizeable fraction of the total force at very early times ($a\sim0.1$), when the Vainshtein radius is still not large enough to screen scales of order $k \sim 1 h\rm{Mpc}^{-1}$. Hence, the fifth force can increase the clustering power until the epoch when the screening starts to become effective on those scales. On the other hand, in the Cubic Galileon model, when the fifth force becomes important, those scales are already within the Vainshtein radius, and therefore, the effects of the fifth force are more effectively screeened. This happens independently of the fact that, at late times, the $G_{\rm{eff}}$ in the Cubic Galileon model is larger than in the models of \cite{Li:2013nua, Wyman:2013jaa}. 

%In the case of the models simulated in \cite{Li:2013nua} and \cite{Wyman:2013jaa}, the linear fifth force represents a sizeable fraction of the total force at very early times, when the density field is still in the linear or mildy nonlinear regime (see Fig.~\ref{H-G-delta} and Fig.~1 of \cite{Wyman:2013jaa}, respectively) and the Vainshtein radius is smaller. Hence, the Vainshtein mechanism has less power in screening the fifth force, since it is not very effective if the matter density fluctuations are small. On the other hand, in the Cubic Galileon model, the fifth force becomes important when the density field is already developed enough so that the Vainshtein mechanism can easily screen the effects of the fifth force. This happens independently of the fact that, at late times, the $G_{\rm{eff}}$ in the Cubic Galileon model is larger than in the models of \cite{Li:2013nua, Wyman:2013jaa}. 

\subsection{Halo mass function}

\begin{figure*}
	\centering
	\includegraphics[scale=0.39]{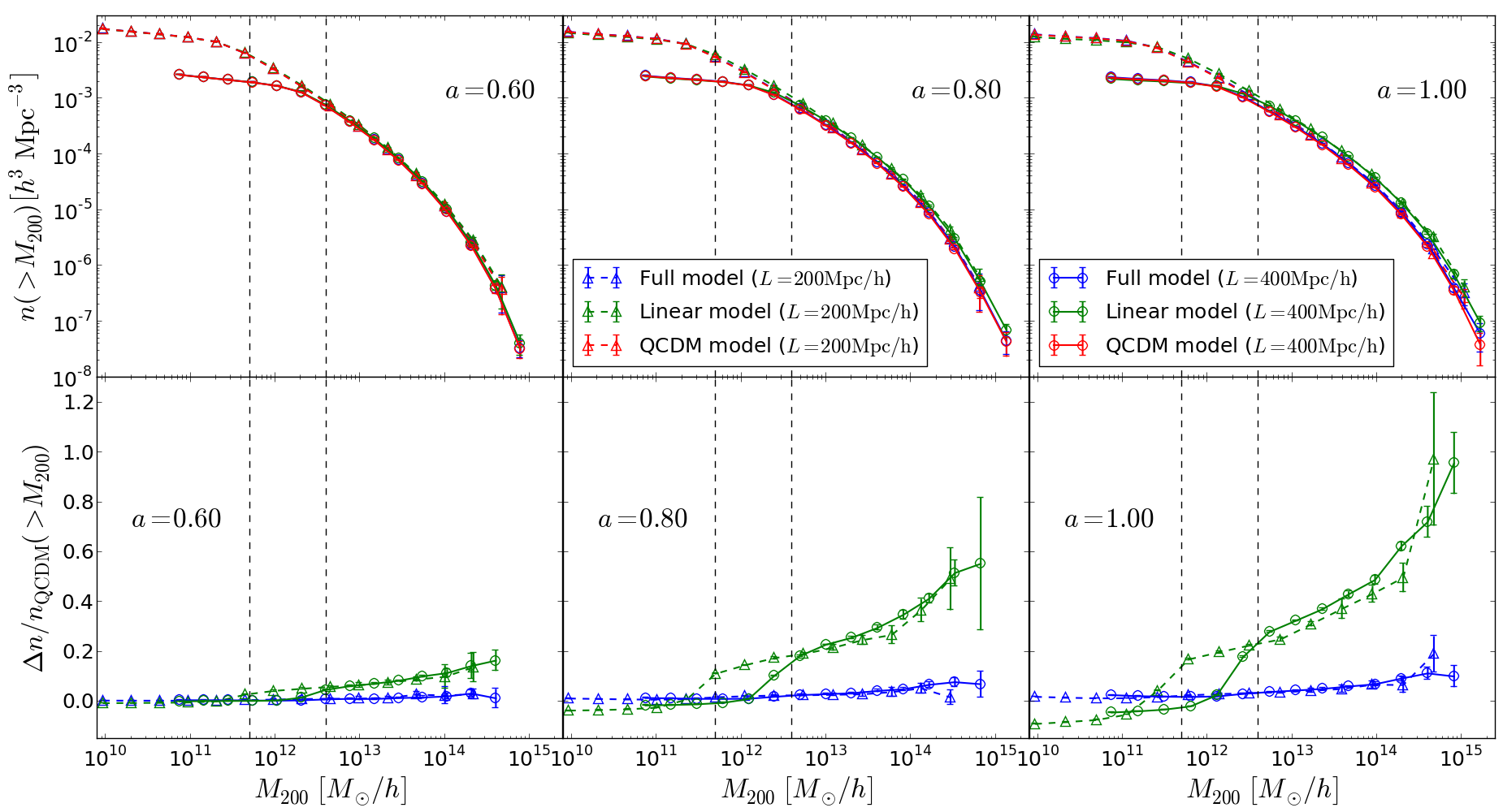}
	\caption{The top panels show the cumulative halo mass function $n\left(>M\right)$ for the Cubic Galileon model (blue) and its linear-force (green) and $\rm{QCDM}$ (red) variants. The vertical black dashed lines give an estimate of the mass resolution threshold $M_{th}$ of the $L = 200\rm{Mpc}/\it{h}$ (dashed lines and open triangles; left vertical dashed line) and $L = 400\rm{Mpc}/\it{h}$ (solid lines and open circles; right vertical dashed line) boxes, where $M_{th} \sim 100m_p = 100 \Omega_m\rho_c L^3/N_p$. The bottom panels show the relative difference to $\rm{QCDM}$, $\Delta n/n_{\rm{QCDM}} = \left(n - n_{\rm{QCDM}}\right)/n_{\rm{QCDM}}$. From left to right the panels correspond to $a = 0.6$, $a = 0.8$ and $a = 1.0$, respectively. The results shown are obtained by averaging over the simulations of the different initial conditions realizations and the errorbars show the standard deviation over these realizations. In the bottom panels we do not show the relative difference at $M_{200} \gtrsim 2\times10^{14}M_{\odot}/h$ and $M_{200} \gtrsim 10^{15}M_{\odot}/h$ for $L = 200\rm{Mpc}/\it{h}$ and $L = 400\rm{Mpc}/\it{h}$, respectively, since for these mass ranges the number of halos is of order unity, which makes the errorbars too large.}
\label{cmf}\end{figure*}

In Fig.~\ref{cmf} we show, in the top panels, the cumulative number density of halos with mass greater than $M$, $n\left(>M\right)$, for the $L = 200\rm{Mpc}/\it{h}$ (dashed lines and open triangles) and $L = 400\rm{Mpc}/\it{h}$ (solid lines and open circles) boxes. The bottom panels show the difference relative to $\rm{QCDM}$, $\Delta n/n_{\rm{QCDM}} = \left(n - n_{\rm{QCDM}}\right)/n_{\rm{QCDM}}$, and from left to right the panels are for three different epochs as in Fig.~\ref{nonlinear-pk}. The two vertical dashed lines give an estimate of the threshold mass below which resolution effects become important in the two simulation boxes, and which we define as $M_{th} \sim 100m_p = 100 \Omega_m\rho_c L^3/N_p$. Figure~\ref{cmf} shows the good agreement between the two boxes results in the range where both of them should not be affected by resolution effects.

To identify the halos in our simulations we use the publicly available phase space halo finder {\tt ROCKSTAR} \cite{Behroozi:2011ju}. The {\tt ROCKSTAR} code is a friends-of-friends (FoF) based halo finder, but it also computes, for each halo, the equivalent mass 

\bq
\label{M200c}M_{200c} = \frac{4\pi}{3}R_{200}^3200\rho_c,
\eq
which is that enclosed in a region of radius $R_{200}$ that corresponds to the spherically averaged density characterized by $\delta = 200\rho_c$. This is the halo mass definition we use in this paper (see \cite{Knebe:2011rx, Knebe:2013xwz} for a comparison and review of halo finding techniques).

The halo mass function results of Fig.~\ref{cmf} show that the screening mechanism is once again very efficient in suppressing the effects of the fifth force. For instance, in the linear force model, one can have an approximately $45\%$ ($35\%$) increase in the number density of halos with masses equal or larger than $10^{14}M_{\odot}/h$ at $a = 1$ ($a = 0.8$). On the other hand, the corresponding increase  in the full model is at most $\sim 10\%$ ($\sim5\%$) at $a = 1$ ($a = 0.8$).

The way the halo abundance responds to the modified force law in the Cubic Galileon model differs considerably from the way it responds, for instance, in the phenomenological model of \cite{Wyman:2013jaa}. In the latter, the authors found that even when the screening is present, the number density of halos with masses of about $10^{15}M_{\odot}/h$ can be $\sim 100\%$ larger, compared to a model without fifth force (see Fig.~2 therein). Such an enhancement is comparable to the one we find for the linear-force model. Once more, the reason behind this discrepancy can be traced back to the different time evolution of the magnitude of the fifth force in the two models. In the Cubic Galileon model, by the time the fifth force becomes non-negligible ($a\sim 0.5$), the length scales relevant for halo formation $\left(k \sim \mathcal{O}(1) h\rm{Mpc}^{-1}\right)$ are already within the Vainshtein radius, and therefore, the fifth force affects them only marginally. But even if there is no screening present, in between $a = 0.5$ and today, the fifth force does not have time to produce an overabundance of halos which is significantly larger than $100\%$ at any mass scale. On the other hand, in the model of \cite{Wyman:2013jaa}, the fifth force becomes important at much earlier times ($a\sim0.1$), when the Vainshtein screening has little impact, and hence, the fifth force has enough time to help enhance structure formation (e.g.~see \cite{Hellwing:2010mrs} for early fifth-force effects on halo abundance). A similar thing happens in simulations of normal-branch DGP models (positive fifth force) \cite{Schmidt:2009sv}.

At the low-mass end of the mass function, although already inside the region where the results may be affected by resolution effects, it seems that there is a trend for the linear-force model simulations to develop a deficiency of low-mass halos with time. This can be explained by the fact that smaller halos in the linear-force model simulations form and virialize earlier, as a consequence of the enhanced gravitational strength. These smaller halos then merge to form larger and more massive halos, which explains why the number of smaller halos decreases with time. A similar result was also found in the simulations of a phenomenological modified gravity model with a Yukawa-like potential fifth force \cite{Hellwing:2011ne, Hellwing:2008qf, Hellwing:2010adp}. Nevertheless, due to the uncertainties related to the resolution effects on these low-mass scales, we remain cautious in deriving a definite conclusion. In future work we plan to address this issue with more detail.

\section{Conclusion}\label{conclusion-section}

Using the {\tt ECOSMOG} code, we have performed the first N-body simulations of Cubic Galileon cosmologies to study the formation of structure in both the linear and nonlinear regimes. This model is a theoretically consistent subset of the most general covariant Galileon theory. As such, it employs the Vainshtein mechanism to screen the modifications of gravity on small scales. The Cubic Galileon model is in many aspects similar to the DGP braneworld model. For instance, the equations solved in the N-body code have the same structure in terms of the spatial derivatives, differing only in the time evolution of the coefficients. An important advantage of the Cubic Galileon, however, is that it can explain the present-day accelerated expansion of the Universe without any ghost-like instabilities, which are known to plague the DGP model.

At the linear level, we have shown that the modified expansion history plays the dominant role compared to the fifth force, in terms of modifying the amplitude of the matter power spectrum. The fifth force, which starts to be important at relatively late times ($a \gtrsim 0.5$), is only responsible for a $5\%$ increase in $\delta^2$ relative to $\Lambda$CDM. This is smaller than the $15\%$ change caused by the modified Hubble expansion rate (c.f. Fig.~(\ref{H-G-delta})).

We derived the equations to be solved by {\tt ECOSMOG} by using the quasi-static approximation, but have identified and circumvented a problem with this procedure. We have shown that in underdense regions, such as voids, there might not be any real solutions for the magnitude of the fifth force (c.f.~Eq.~(\ref{discriminant})). This is also a problem in the DGP model which, however, is not as bad as in the Galileon model due to the different time evolution of the coefficients of the spatial derivatives (c.f. Fig.~\ref{beta-rv}). In our simulations, to overcome this problem, we have adopted the simple strategy of fixing to zero the values of the square roots that would have become imaginary. This solution, which may appear artificial, is not expected to have a significant impact on the small scale nonlinear power spectrum. The reason we think this is the case is because the power spectrum on small scales is dominated by high density regions where the problem does not appear. Moreover, we have also pointed out that it is not clear that our solution to this problem is making the fifth force calculation inaccurate. We noted that the problem appears in regions where the quasi-static limit is not expected to be a good approximation, and therefore, the calculation would still be inaccurate even if the solutions would be kept real. To clarify the true impact of our solution to the problem, and also of the quasi-static approximation, one would need to simulate the equations derived without any assumptions. This investigation is much more challenging and we leave it for future work.

Our results show that the Vainshtein mechanism can work very effectively to suppress the effects of the fifth force. At scales of the order of $k \sim 1 h\rm{Mpc}^{-1}$, the unscreened fifth force, which is roughly twice as strong as normal gravity at the present time (c.f.~Fig.~\ref{H-G-delta}), leads to an increase of the clustering power of $\sim 300\%$ at $a = 1$. On the other hand, if one turns on the screening, the fifth force does not increase the power, at any scale or time, by more than $\sim 8\%$. To our knowledge, of all the models with the Vainshtein mechanism whose nonlinear structure formation has been studied with N-body simulations so far \cite{Schmidt:2009sg, Schmidt:2009sv, Chan:2009ew, Khoury:2009tk, Li:2013nua, Wyman:2013jaa}, the Cubic Galileon is the model in which the screening has been found to be the most effective. This is due to the (unscreened) fifth force becoming relevant only at late times, when the Vainshtein radius of individual matter sources is sufficiently large for the screening mechanism to be very efficiently implemented on scales $k \gtrsim 0.1 h\rm{Mpc}^{-1}$.

The same effectiveness of the Vainshtein mechanism was found in the results for the halo mass function. For instance, at $a = 1$, the number density of halos with masses $\gtrsim 10^{14}M_{\odot}/h$ is $\sim 45\%$ larger in the linear force model than in the $\rm{QCDM}$ model. The same difference, however, becomes only less than about $10\%$, if the screening is implemented (c.f. Fig.~\ref{cmf}). As before, the success of the Vainshtein mechanism in screening the fifth force is related to the late time deviation from unity of the ratio $G_{\rm{eff}}/G$. Our results also show a deficiency in the number of low-mass halos in the linear-force model simulations. However, the mass of such small halos falls below the mass resolution of our simulations, and therefore, we leave for future work a better investigation of this result.

Our work shows that the presence of the screening mechanism does not change the predictions from linear perturbation theory on large scales $k \lesssim 0.1h \rm{Mpc}^{-1}$. As a result, the $\sim 20\%$ ($a = 1$, c.f.~Fig.~\ref{H-G-delta}) increase of the amplitude of the power spectrum on these scales compared to $\Lambda$CDM anticipates that galaxy clustering measurements can place very strong constraints on the Cubic Galileon model. In addition, we have also shown that the Cubic Galileon model fails to provide a good fit to the low-$l$ ($l \lesssim 50$) region of the CMB temperature power spectrum, although it fits the higher multipoles as well as $\Lambda$CDM. The best-fitting Cubic Galileon model to a combination of data from the CMB, SNIa and BAO, has too much ISW power on the largest angular scales to be compatible with the observations (c.f. Fig.~(\ref{Cl})). With respect to the general Galileon model, which is able to provide a good fit to the CMB data \cite{Barreira:2013jma}, the Cubic Galileon model is disfavoured by $\chi^2_{\rm{Cubic}} - \chi^2_{\rm{general}} \approx 16.5$. This hints that the success of the Galileon model may require all of the Lagrangian terms to be at play, rather than just a subset of them. Such an investigation is left for future work. For now, the conclusion is that, as it stands, the Cubic Galileon model is already in tension with the current data. This fact, combined with the stability and strong coupling problems that have been associated with the higher order Galileon Lagrangian terms \cite{PhysRevD.79.064036, Burrage:2012ja}, may also represent a serious problem to the general Galileon gravity model.

We stress that although the Cubic Galileon model is in some tension with the current data, it is still valuable to investigate the impact that different modified gravity models can have on nonlinear structure formation. This allows for a better understanding of the ways to distinguish between different modified gravity theories and also for a more robust interpretation of the data from current and future large scale structure surveys, such as the forthcoming BigBOSS \cite{Schelgel:2011zz} and Euclid \cite{Laureijs:2011gra, Amendola:2012ys} missions.  In the future, we plan to continue to study more large scale structure properties of this model, in particular, halo bias, redshift space distortions and the BAO scale feature.

\begin{acknowledgments}

AB is supported by FCT-Portugal through grant SFRH/BD/75791/2011. BL is supported by the Royal Astronomical Society and Durham University. WAH acknowledges the support received from the Polish National Science Center through grant no. DEC-2011/01/D/ST9/01960 and ERC Advanced Investigator grant of C.~S.~Frenk, COSMIWAY. This work has been partially supported by the European Union FP7  ITN INVISIBLES (Marie Curie Actions, PITN- GA-2011- 289442) and STFC. The N-body simulations presented in this paper were run on the ICC Cosmology Machine, which is part of the DiRAC Facility jointly funded by STFC, the Large Facilities Capital Fund of BIS, and Durham University.

\end{acknowledgments}

\bibliography{cubic-simulations.bib}

\end{document}